\documentclass{ieeeaccess}

\usepackage[hyphens]{url}
\usepackage[hidelinks]{hyperref}
\hypersetup{breaklinks=true}
\usepackage{cite}
\usepackage{amsmath,amssymb,amsfonts}
\usepackage{algorithm}
\usepackage{algorithmic}
\usepackage{graphicx}
\usepackage{textcomp}
\usepackage{afterpage}
\usepackage{array}
\usepackage{supertabular}
\usepackage{multirow}
\usepackage{multicol}
\usepackage{multirow}
\usepackage{afterpage}
\usepackage{siunitx}
\usepackage{color}
\usepackage{subfigure}
\usepackage{makecell}
\usepackage{txfonts}
\usepackage{rotating}
\usepackage{romannum}
\usepackage{tipa}
\usepackage{bm}
\usepackage{amssymb}
\usepackage{comment}

\def\BibTeX{{\rm B\kern-.05em{\sc i\kern-.025em b}\kern-.08em
    T\kern-.1667em\lower.7ex\hbox{E}\kern-.125emX}}
\begin{document}
\history{Date of publication xxxx 00, 0000, date of current version xxxx 00, 0000.}
\doi{10.1109/ACCESS.2023.DOI}

\title{IR-UWB Radar-Based Contactless Silent Speech Recognition of Vowels, Consonants, Words, and Phrases}

\author{\uppercase{Sunghwa Lee}\authorrefmark{1}, 
\uppercase{Younghoon Shin}\authorrefmark{1}, 
\uppercase{Myungjong Kim}\authorrefmark{2}, 
\uppercase{and Jiwon Seo}\authorrefmark{1, 3}, \IEEEmembership{Member, IEEE}}

\address[1]{School of Integrated Technology, Yonsei University, Incheon 21983, Republic of Korea}
\address[2]{NVIDIA Corporation, Santa Clara, CA 95051, United States}
\address[3]{Department of Convergence IT Engineering, Pohang University of Science and Technology, Pohang 37673, Republic of Korea}

\tfootnote{
This work was supported by the National Research Foundation of Korea (NRF) grant funded by the Korea government (MSIT) (NRF-2021R1F1A1062958).
}

\markboth
{Lee \textit{et al.}: IR-UWB Radar-Based Contactless Silent Speech Recognition of Vowels, Consonants, Words, and Phrases}
{Lee \textit{et al.}: IR-UWB Radar-Based Contactless Silent Speech Recognition of Vowels, Consonants, Words, and Phrases}

\corresp{Corresponding author: Jiwon Seo (jiwon.seo@yonsei.ac.kr)}

\begin{abstract}
Several sensing techniques have been proposed for silent speech recognition (SSR); however, many of these methods require invasive processes or sensor attachment to the skin using adhesive tape or glue, rendering them unsuitable for frequent use in daily life. 
By contrast, impulse radio ultra-wideband (IR-UWB) radar can operate without physical contact with users' articulators and related body parts, offering several advantages for SSR. 
These advantages include high range resolution, high penetrability, low power consumption, robustness to external light or sound interference, and the ability to be embedded in space-constrained handheld devices.
This study demonstrated IR-UWB radar-based contactless SSR using four types of speech stimuli (vowels, consonants, words, and phrases). 
To achieve this, a novel speech feature extraction algorithm specifically designed for IR-UWB radar-based SSR is proposed. 
Each speech stimulus is recognized by applying a classification algorithm to the extracted speech features. 
Two different algorithms, multidimensional dynamic time warping (MD-DTW) and deep neural network–-hidden Markov model (DNN--HMM), were compared for the classification task. 
Additionally, a favorable radar antenna position, either in front of the user's lips or below the user's chin, was determined to achieve higher recognition accuracy. 
Experimental results demonstrated the efficacy of the proposed speech feature extraction algorithm combined with DNN--HMM for classifying vowels, consonants, words, and phrases.
Notably, this study represents the first demonstration of phoneme-level SSR using contactless radar.
\end{abstract}


\titlepgskip=-15pt

\maketitle

\section{Introduction}
\label{sec:Intro}

Speech is an attractive input modality for human--computer interaction owing to its convenience and efficiency.
Automatic speech recognition (ASR) technology has achieved high accuracy and robustness for deployment in commercial products.
For example, several recent commercial smart devices have adopted digital voice assistance services, such as Amazon Alexa, Apple Siri, Microsoft Cortana, Google Assistant, and Samsung Bixby.
However, audio-based ASR has certain limitations.
In noisy environments (e.g., concert halls), performance may degrade significantly.
Furthermore, its usage is limited in places where silence has to be maintained (e.g., libraries) or in situations where confidential speech communication is needed (e.g., military operations).
Finally, it is not usable by people who have lost their voices because of reasons (e.g., laryngectomy).

Myriad acoustic and nonacoustic biosignals can be captured during speech production.
Silent speech recognition (SSR) is a technology that converts the nonacoustic speech-related biosignals captured from body parts, such as the brain, muscles, and organs, into text.
Because SSR does not require auditory information, it can be a standalone solution for speech-based human--computer interactions in situations where ASR cannot be applied, or it can be used together with ASR to enhance performance.

Various sensing techniques have been proposed for capturing nonacoustic speech-related biosignals \cite{Denby2010, Schultz2017, Gonzalez2020:SSIreview, Lee2021:SSIreview}.
Among these techniques, electromagnetic articulography (EMA) \cite{Schonle1987, Fagan2008, Heracleous2011:EMA, Bocquelet2016:EMA, Wang2016:EMA, Kim2017}, permanent magnetic articulography \cite{Gilbert2010, Gonzalez2016}, and electropalatography \cite{Li2019:EPG, Woo2021, Kimura2022:EPG} involve the placement of sensors inside the oral cavity.
The tongue is one of the primary articulators; however, capturing tongue motion is challenging because the tongue is inside the oral cavity.
These techniques are advantageous for capturing tongue motion.
However, they are not appropriate for daily use because of cumbersome sensor attachment procedures and inconvenience to users.

Surface electromyography \cite{Wand2011, Meltzner2017}, vision \cite{Cetingul2006:image, Yau2006:image, su2018:image, Sheng2021:image, Lopez2022:image, Haq2022:image}, ultrasound imaging \cite{Hueber2007:US, Gosztolya2019:US}, and radar \cite{Birkholz2018:radar, Digehsara2022, Shin16:Towards, Wen2020, Ferreira2022:radar, Zeng2023} are techniques for capturing nonacoustic speech-related biosignals without the need to place sensors inside the oral cavity.
Although these techniques are more convenient than the aforementioned ones, they have some shortcomings.

Surface electromyography, which can be used to detect the activities of facial muscles during speech production using electrodes, intrinsically suffers from high signal variability between speech sessions owing to variations in sensor placement \cite{Prorokovic2019:sEMG}.
Moreover, their usability is limited because they require the attachment of electrodes to the skin.

Although images are a highly accessible modality because many devices used in daily life, such as smartphones and laptops, are equipped with image sensors, vision techniques have intrinsic performance limitations because they cannot detect invisible articulators.
Furthermore, they pose privacy concerns, and their performance depends heavily on the light conditions.

Ultrasound imaging, which enables intraoral scanning, has problems of limited quality, caused partly by the presence of speckle noise and loss of signal from the part of the tongue that is not orthogonal to the ultrasound beam \cite{Schultz2017}.
Denby \textit{et al.} \cite{Denby2011} suggested a lightweight helmet combining an infrared camera and an ultrasound device to improve the SSR performance. 
Infrared cameras and ultrasound devices are efficient in detecting lip and tongue motion, respectively, which are the main articulators for speech production.
However, this approach requires wearing a helmet, which can deteriorate the usability.

\begin{table*}
\renewcommand{\arraystretch}{1.3}
\setlength{\tabcolsep}{2.5em}
\centering
\caption{Comparison of radar-based silent speech recognition studies.}
\label{tab:RadarStudies}
\begin{tabular}{ccccc}
\hline \hline
\textbf{Reference}                                 & \textbf{Year}       & \textbf{Data acquisition mode} & \textbf{Corpus} \\ \hline
Birkholz \textit{et al.} \cite{Birkholz2018:radar} & 2018                & Contact                        & 25 phonemes \\ \hline
Digehsara \textit{et al.} \cite{Digehsara2022}     & 2022                & Contact                        & 40 words  \\ \hline
Shin and Seo \cite{Shin16:Towards}                 & 2016                & Contactless                    & 10 words and 5 vowels \\ \hline
Wen \textit{et al.} \cite{Wen2020}   & 2020  & Contactless      & N/A \\ \hline
Ferreira \textit{et al.} \cite{Ferreira2022:radar} & 2022                & Contactless                    & 13 words  \\ \hline
Zeng \textit{et al.} \cite{Zeng2023} & 2023  & Contactless      & 1000 sentences \\ \hline
Proposed                                           & 2023                & Contactless                    & 8 vowels, 11 consonants, 25 words, and 12 phrases  \\ \hline \hline
\end{tabular}
\end{table*}

Radar is a promising technique for SSR because it works through occluding materials and is unaffected by external light or sound conditions.
However, extracting effective speech features from radar signals is challenging, because raw radar signals are very complex to interpret. 
Furthermore, radar signals may contain motion information of non-speech sources other than the articulators.
Therefore, most studies on radar-based SSR have not demonstrated results beyond the word recognition level, as summarized in Table \ref{tab:RadarStudies}.

An exception is the work of Birkholz \textit{et al.} \cite{Birkholz2018:radar}, who used an ultra-wideband (UWB) radar to recognize 25 German phonemes.
They attached two antennas to the cheek and below the chin of the speakers using adhesive tape, which allowed them to capture speech movements more effectively.
However, these skin-attached antennas are not suitable for daily use because they are inconvenient to users.
Digehsara \textit{et al.} \cite{Digehsara2022} recently proposed a wearable headset for SSR equipped with a stepped-frequency continuous-wave radar and two antennas on the left and right cheeks; however, they tested it for word recognition only.

A contactless SSR system is much more desirable because its usability is superior to that of a contact SSR system, which requires a helmet or a skin-attached antenna. 
The usability of a contactless SSR system embedded in smart devices is discussed in Section \ref{sec:FutureSSR}. 
The following studies utilized radar to implement a contactless SSR solution.

Shin and Seo \cite{Shin16:Towards} demonstrated the recognition of 10 isolated words and 5 vowels using a bistatic impulse radio ultra-wideband (IR-UWB) radar.
To implement an IR-UWB radar-based contactless SSR system, Shin and Seo \cite{Shin16:Towards} proposed a method to extract the distance and correlation amplitude from raw radar measurements as speech features for SSR.
However, the five vowels they used (/\textscripta/, /\textipa{\ae}/, /\textipa{i}/, /\textipa{o}/, and /\textipa{u}/) are highly distinguishable. 
Thus, recognizing them does not pose a greater challenge than recognizing words.
The authors also stated that their proposed features are insufficient to enable phoneme-level recognition.

Wen \textit{et al.} \cite{Wen2020} captured speech movements with a customized radar sensor capable of dual frequency-modulated continuous-wave (FMCW) and continuous-wave (CW) modes.
They demonstrated that the displacement and spectrum patterns, obtained using their algorithm to resolve phase ambiguity caused by the nonlinear phase modulation of their radar system, for several word and sentence commands were distinct.
However, they did not conduct a speech recognition experiment. 

Ferreira \textit{et al.} \cite{Ferreira2022:radar} performed a speech recognition task with 13 isolated European Portuguese words using an FMCW radar.
They utilized velocity dispersion data as speech features and successfully demonstrated that these features were capable of classifying distinguishable words.
However, these accomplishments have not yet been extended to phoneme recognition.

Zeng \textit{et al.} \cite{Zeng2023} conducted the recognition of individual words within 1000 sentences of everyday conversation using an FMCW radar. 
They introduced a novel signal processing pipeline that sequentially localizes articulatory zones, removes low-frequency noise, and extracts short-time Fourier transform-based speech features. 
Additionally, they designed an end-to-end deep neural network to recognize words from sentences based on these speech features. 
While their achievement in sentence-level word recognition is promising, they did not provide phoneme-level recognition results or analysis.
This omission limits a comprehensive understanding of the potential or limitations of their SSR system.

Compared with conventional radars, the IR-UWB radar used in our study has promising properties for deployment in SSR. 
It has a higher performance potential than conventional radars owing to its higher range resolution \cite{Fontana2004, Skaria2020}, better signal penetrability \cite{Kim2018}, and lower power consumption \cite{Fernandes2010}.
For instance, Wang \textit{et al.} \cite{Wang2020} demonstrated the superior accuracy and signal-to-noise ratio of IR-UWB radar over FMCW radar in nearly all comparison scenarios for contactless vital sign monitoring, which measures respiration rate and heart rate.
However, as introduced earlier, fewer studies have focused on IR-UWB radar-based SSR compared to FMCW radar-based SSR up to the present date.

Leveraging the high-performance potential of IR-UWB radar, our study aims to accomplish phoneme-level speech recognition. 
To transcribe speakers' real-life conversations into text, SSR should possess the capability to recognize phonemes themselves rather than words composed of multiple phonemes. 
However, to the best of our knowledge, contactless radar-based SSR studies in the literature have not yet demonstrated phoneme-level recognition.

One of the main challenges in contactless phoneme-level SSR using radar is defining and extracting appropriate speech features from raw radar data.  
Given the different working mechanisms of FMCW and IR-UWB radar systems \cite{Wang2020}, the signal processing algorithms employed in FMCW radar-based SSR cannot be directly transferred to IR-UWB radar-based SSR.
In this study, we proposed a new speech feature extraction algorithm that can capture the necessary articulatory movements from IR-UWB radar data for phoneme-level SSR.
Silent speech recognition of phonemes (8 vowels and 11 consonants), 25 words, and 12 phrases was demonstrated using a contactless IR-UWB radar.
Furthermore, we applied classification algorithms to the extracted speech features to analyze how the performance varied according to the choice of the classification algorithm.
This was the first feasibility test for silent phoneme-level recognition, including both vowels and consonants, using a contactless radar platform.

\begin{figure*}
\centerline{\includegraphics[width=0.90\linewidth]{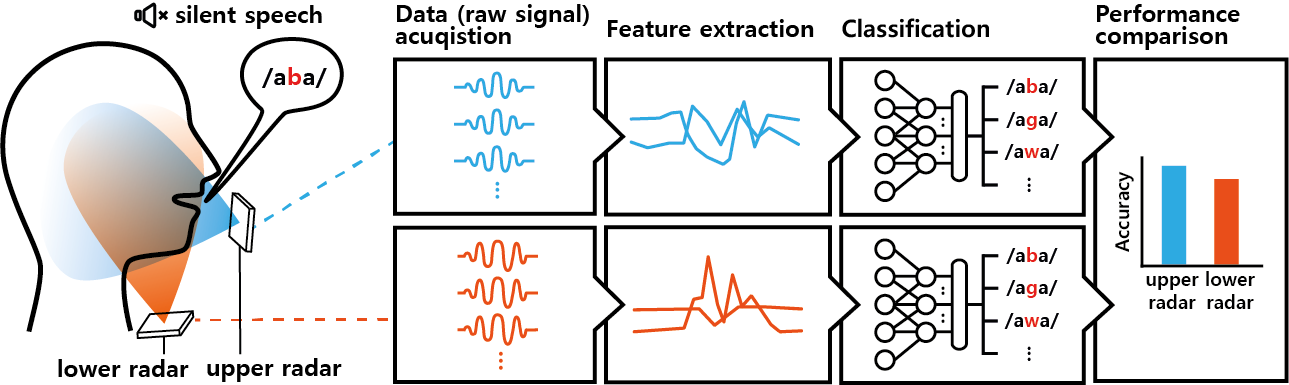}}
\caption{Overall schematic of our approach for the contactless IR-UWB radar-based SSR study.
Articulatory movements are captured simultaneously by two radars, with one antenna positioned at the upper radar position and the other at the lower radar position, to investigate the optimal antenna position for SSR.
Feature extraction and classification algorithms are independently applied to the upper and lower radar signals.
}
\label{fig:Overview}
\end{figure*}

Another research question in this study concerns the favorable direction of radar signals for SSR.
Positioning the radar antenna in front of the lips (i.e., the upper radar case in Fig. \ref{fig:Overview}) is a promising strategy, because the movements of the lips and tongue can be captured together.
However, positioning the radar antenna under the speaker's chin (i.e., the lower radar case in Fig. \ref{fig:Overview}) is another possible strategy, because the movements of the tongue and its related body parts can be detected by the lower radar \cite{Lee22:lower}.
To answer this research question, we positioned two radar antennas, one each at the front of the lips and under the chin, during the experiments and captured articulatory movements using these two radars simultaneously. 
We then compared the SSR performance of the two cases.

The contributions of this study are summarized as follows:
\begin{itemize}
    \item The IR-UWB radar-based contactless SSR task was performed using three speech-unit levels: phonemes, words, and phrases. 
    This study included the first contactless radar-based SSR of phonemes, including both vowels and consonants.
    \item A novel speech feature extraction algorithm was proposed to improve the performance of IR-UWB radar-based SSR.
    \item Two classification algorithms were implemented and applied to the recognition task along with the proposed feature extraction algorithm.
    \item An analysis of the choice between two radar positions---in front of the speaker's lips or under the speaker's chin---that is more favorable for a higher performance of the IR-UWB radar-based SSR was conducted.
\end{itemize}

The remainder of this paper is organized as follows: 
Section \ref{sec:Principles} explains the basic working principles of IR-UWB radar-based SSR, and Section \ref{sec:DataAcquisition} describes our testbed and data acquisition procedure. 
Section \ref{sec:Method} presents the details of the proposed speech feature extraction algorithm along with classification algorithms. 
After discussing the performance and effectiveness of the proposed method in Section \ref{sec:ResultsEval}, conclusions are presented in Section \ref{sec:Conclusion}.

\section{Principles of IR-UWB Radar-Based SSR}
\label{sec:Principles}

Radar can be broadly classified into CW radar and pulse radar. 
CW radar transmits a continuous wave with a constant frequency, allowing for the measurement of target velocity through Doppler shift analysis. 
However, it faces challenges in independently determining target range.

An FMCW radar, a variation of CW radar, emits a continuous wave with a frequency that changes over time. 
It can measure both the target range and velocity by analyzing the beat frequency between the transmitted and received signals. 
When FMCW radar operates in the frequency range of 30 to 300 GHz, it is commonly referred to as mmWave radar because the wavelength of the radio waves is in the millimeter range. 
Specifically, the FMCW radar employed in the contactless radar-based SSR studies of \cite{Wen2020} and \cite{Zeng2023}, introduced in Section \ref{sec:Intro}, corresponds to mmWave radar.

In contrast, pulse radar transmits short pulses, enabling the measurement of target range and velocity by analyzing the amplitude and time of flight of these pulses. 
When the pulse radar transmits extremely short pulses with a fractional bandwidth larger than 25\%, it is categorized as IR-UWB radar. 
The fractional bandwidth is defined as the ratio of the bandwidth to the center frequency.

FMCW radar utilizes continuous wave signals, whereas IR-UWB radar employs pulse signals for its functionality. 
Thus, the signal processing algorithm used in the FMCW radar-based SSR studies of \cite{Wen2020, Ferreira2022:radar, Zeng2023} cannot be directly applied to our IR-UWB radar-based SSR study.

In the remainder of this section, we explain the detailed operational mechanism of the IR-UWB radar used in this study.
The radar's transmit (TX) antenna emits pulses that travel through air and are subsequently reflected by targets.
The reflected pulses are then received by the receive (RX) antenna of the radar and merged into a data ``frame'' after employing the IR-UWB radar manufacturer's proprietary normalization method.

\begin{figure}
\centerline{\includegraphics[width=1\linewidth]{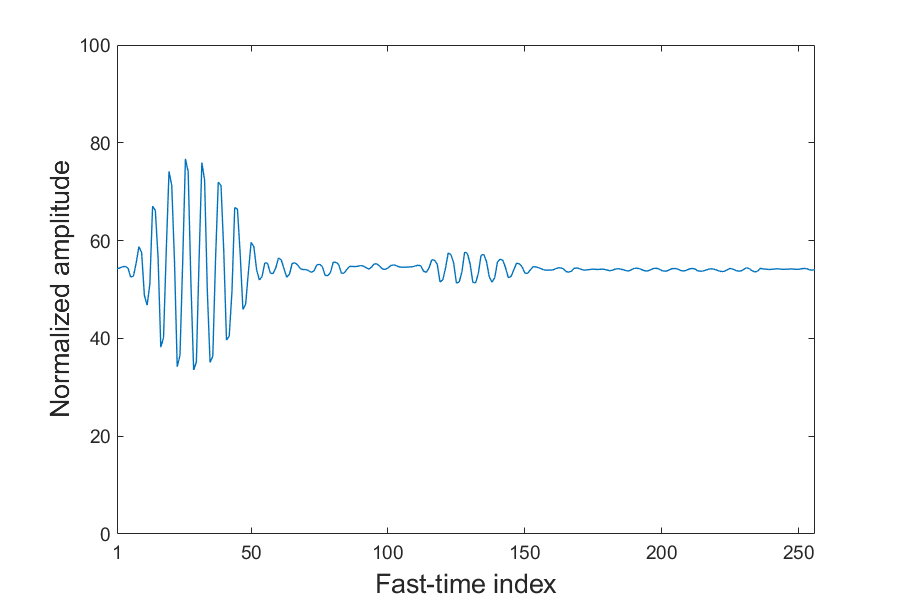}}
\caption{A single data frame captured using the IR-UWB radar. 
This data frame was captured 1.5 s after the onset of the pronunciation, while the sentence ``How are you doing?'' was silently pronounced.}
\label{fig:FrameExample}
\end{figure}

Fig. \ref{fig:FrameExample} illustrates an example of a single data frame while the sentence ``How are you doing?'' was silently pronounced. 
This data frame was captured 1.5 s after the onset of the pronunciation. 
The radar's detection range was set to cover distances of up to 1 m. 
Each ``fast-time'' index, ranging from 1 to 256, corresponds to a specific distance within this 1 m range. 
For example, the fast-time index of 26 corresponds to $\frac{26}{256} \times 1 \, \text{[m]} = 0.10 \, \text{[m]}$ from the radar antenna. 
The normalized signal amplitude, which ranges from 0 to 100, represents the strength of the reflection at the corresponding distance (i.e., the fast-time index). 
In Fig. \ref{fig:FrameExample}, the highest amplitude occurs at a fast-time index of 26, indicating that the target responsible for the most intense pulse reflection is approximately located 0.1 m from the radar antenna. 
The amplitudes of the reflected pulses demonstrate diverse time delays and shapes, which correspond to the characteristics of the targets, including distance, shape, and angle.

\begin{figure}
  \centering
  \includegraphics[width=1\linewidth]{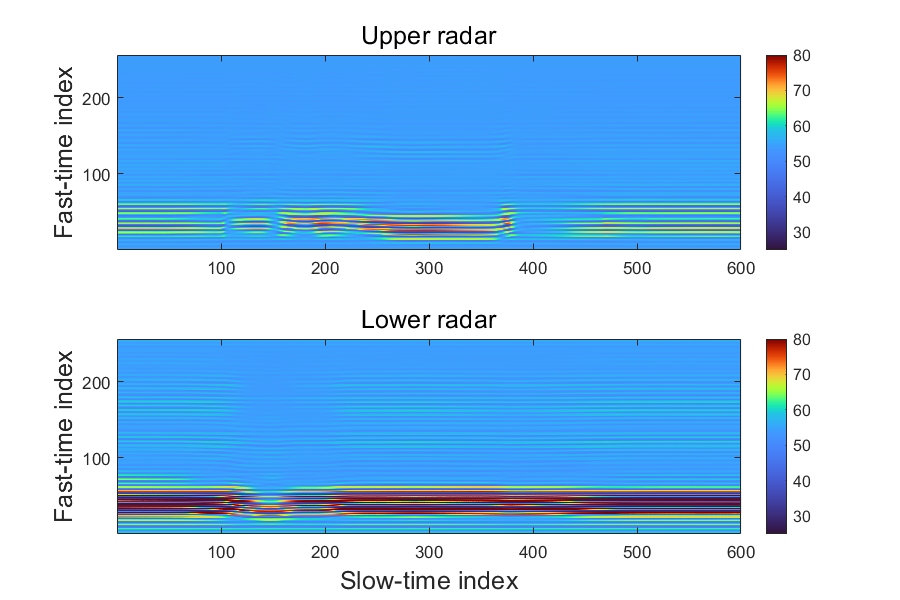}
  \caption{Raw frame sets that capture the articulatory movements for silently pronouncing ``How are you doing?''. 
  Two frame sets were acquired by the upper and lower radars in Fig.\ref{fig:Overview}.
  The slow-time and fast-time indices indicate the time and distance, respectively. 
  The color bar represents the normalized signal amplitude.}
  \label{fig:RawFrameset}
\end{figure}

The IR-UWB radar used in this study continuously acquired data frames at an approximate rate of 200 frames per second. 
These acquired data frames encompass information regarding the articulatory movements involved in speech production. 
A ``frame set'' is formed by combining $M$ one-dimensional data frames, each with a length of 256, resulting in a frame set with dimensions of $256$-by-$M$. 
Fig. \ref{fig:RawFrameset} shows two frame sets, each comprising $M = 600$ (equivalent to 3 s of data), from the upper and lower radars. 
The colors in the visualization correspond to the normalized signal amplitudes. 
Typically, the row index of the frame set is referred to as the ``fast-time'' index, representing the distance to the target, while the column index is known as the ``slow-time'' index and indicates the reception time of the corresponding data frame. 
For instance, the data frame depicted in Fig. \ref{fig:FrameExample} corresponds to a slow-time index of 300, signifying that the data frame was acquired at approximately ${300 \, \text{[frames]}} / {200 \, \text{[frames per second]}} = 1.5 \, \text{[s]}$ after the onset of the pronunciation. 
That is, the data frame illustrated in Fig. \ref{fig:FrameExample} represents a two-dimensional slice of the three-dimensional data shown in Fig. \ref{fig:RawFrameset} (top) at a slow-time index of 300.

\begin{figure}
  \centering
  \includegraphics[width=1\linewidth]{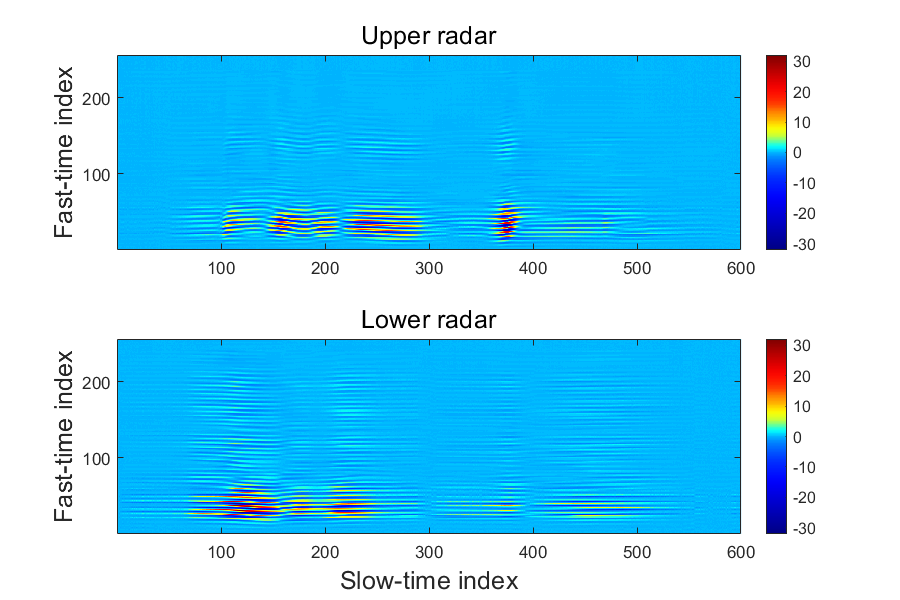}
  \caption{Clutter-reduced frame sets that capture the articulatory movements for silently pronouncing ``How are you doing?''. 
  Two frame sets were acquired by the upper and lower radars in Fig.\ref{fig:Overview}.
  The slow-time and fast-time indices indicate the time and distance, respectively.
  The color bar represents the normalized signal amplitude.}
  \label{fig:ClutterreducedFrameset}
\end{figure}

In radar applications, the term ``clutter'' refers to undesired signals reflected from stationary or slowly moving objects in the surrounding environment. 
The clutter-reduced frame sets shown in Fig. \ref{fig:ClutterreducedFrameset} are obtained by subtracting the clutter from the raw frame sets illustrated in Fig. \ref{fig:RawFrameset}. 
In this study, the clutter was estimated using the loopback filter \cite{Leem2020} and subtracted from the raw frame set using the following equations:
\begin{equation}
  c_{m}[n] = \alpha c_{m-1}[n] + (1 - \alpha) r_{m}[n]
  \label{eq:LBF_clutter}
\end{equation}
\begin{equation}
  y_{m}[n] = r_{m}[n] - c_{m}[n]
  \label{eq:clutter_reduction}
\end{equation}
Here, $c_{m}[n]$, $r_{m}[n]$, and $y_{m}[n]$ represent the clutter amplitude, received signal amplitude, and clutter-reduced signal amplitude, respectively, at slow-time index $m$ and fast-time index $n$. 
The value of $\alpha$ was set to 0.95 in this study. 
It is important to note that $r_{m}[n]$ and $y_{m}[n]$, where $m$ ranges from 1 to $M$ and $n$ ranges from 1 to 256, form matrices that represent the raw and clutter-reduced frame sets, respectively.

The raw and clutter-reduced frame sets capturing the articulatory movements of a participant while silently pronouncing ``How are you doing?'', are shown in Figs. \ref{fig:RawFrameset} and \ref{fig:ClutterreducedFrameset}, respectively. 
In both the raw and clutter-reduced frame sets, the normalized signal amplitude at each fast-time index (i.e., distance) varies as the slow-time index (i.e., time) changes. 
This indicates that the distance between articulators and radar varies over time during pronunciation. 
Notably, stationary signals in the raw frame sets are significantly reduced in the clutter-reduced frame sets. 
Consequently, the changes in amplitude are more pronounced in Fig. \ref{fig:ClutterreducedFrameset} than in Fig. \ref{fig:RawFrameset}.

While we acknowledge the variation in radar data observed in Figs. \ref{fig:RawFrameset} or \ref{fig:ClutterreducedFrameset} during silent pronunciation, it remains challenging to define and extract suitable speech features from this data that facilitate the recognition of phonemes within the silently uttered sentence ``How are you doing?''.

\section{Data Acquisition}
\label{sec:DataAcquisition}

Before elucidating the proposed methods, this section explains the speech stimuli, hardware testbed, and test procedure.

\subsection{Speech Stimuli and Participants}
\label{sec:Participants}

All speech stimuli (8 vowels, 11 consonants, 25 words, and 12 phrases) used in this study are based on \cite{Wang2016:EMA}.
More precisely, the vowels and consonants correspond to 8 consonant-vowel-consonant (CVC) and 11 vowel-consonant-vowel (VCV) syllables, respectively.
Pronouncing these CVC and VCV syllables involves diverse mechanisms for articulating English vowels and consonants.
The 25 words used for the isolated word recognition were designed to be phonetically balanced.
Finally, 12 short phrases were used for phrase classification.
These phrases are often used in augmentative and alternative communication devices \cite{Wang2012}.
Detailed information on the pronunciation list used in this study can be found in \cite{Wang2016:EMA}.
The comprehensive pronunciation list is as follows.

\begin{itemize}
  \item 8 CVC syllables: /\textipa{b\textscripta{b}}/, /\textipa{bib}/, /\textipa{beb}/, /\textipa{b\ae{b}}/, /\textipa{b\textturnv{b}}/, /\textipa{b\textopeno{b}}/, /\textipa{bob}/ and /\textipa{bub}/ 
  \item 11 VCV syllables: /\textipa{\textscripta{b}\textscripta}/, /\textscripta\textscriptg\textscripta/, /\textipa{\textscripta{w}\textscripta/}, /\textipa{\textscripta{v}\textscripta}/, /\textipa{\textscripta{d}\textscripta}/, /\textipa{\textscripta{z}\textscripta/}, /\textipa{\textscripta{l}\textscripta}/, /\textipa{\textscripta{r}\textscripta/}, /\textscripta\textyogh\textscripta/, /\textscripta\textdyoghlig\textscripta/, and /\textipa{\textscripta{j}\textscripta}/
  \item 25 words: job, need, charge, hit, blush, snuff, log, nut, frog, gloss, start, moose, trash, awe, pick, bud, mute, them, fate, tang, corpse, rap, vast, dab, and ways
  \item 12 phrases: ``How are you doing?,'' ``I am fine,'' ``I need help,'' ``That is perfect,'' ``Do you understand me?,'' ``Right,'' ``Hello!,'' ``Why not?,'' ``Please repeat that,'' ``Good-bye,'' ``I don't know,'' and ``What happened?''
\end{itemize}

Twenty participants (13 males and 7 females), aged between 20 and 28, were recruited for the experiment. 
Four participants (two males and two females) were native speakers of American English and pronounced 8 CVC syllables, 11 VCV syllables, 25 words, and 12 phrases. 
The remaining 16 participants, native speakers of Korean with at least 13 years of English education, pronounced 8 CVC and 11 VCV syllables. 
Consequently, phoneme recognition tasks, the primary focus of our study and more challenging than word or sentence recognition, were evaluated with data from all 20 participants. 
All participants repeated each speech stimulus 20 times without vocalization.
They were instructed to articulate each speech stimulus clearly and maintain a stationary head position throughout the data acquisition period.

\subsection{Hardware Testbed}

\begin{figure}
\centering
\centerline{\includegraphics[width=0.8\linewidth]{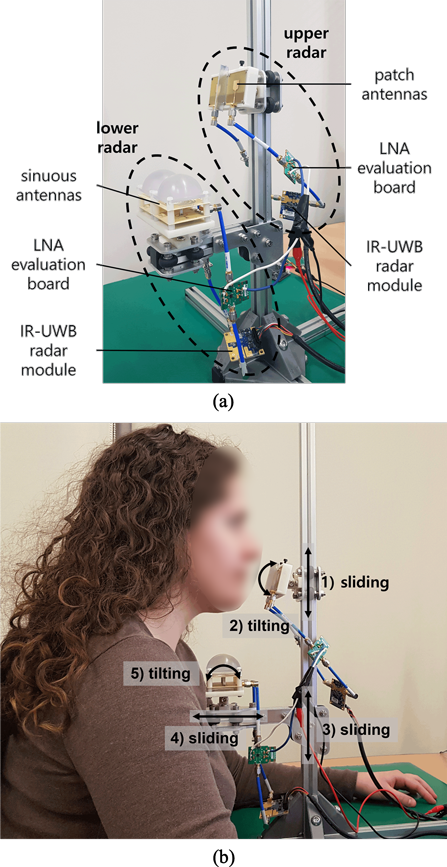}}
\caption{(a) Components and (b) functionalities of the hardware testbed.}
\label{fig:Testbed}
\end{figure}

The IR-UWB radar operates based on pulse modulation, involving the transmission of impulses (extremely short pulses) without a continuous carrier wave \cite{Fontana2004}.
The IR-UWB radar modules used in this study (NVA-R661, Novelda) emit impulses with a frequency range of 6 to 10.2 GHz and provide a 4 mm distance resolution for received signals.

As illustrated in Fig. \ref{fig:Testbed}(a), we constructed a hardware testbed to accurately capture radar data reflecting the articulatory movements of users. 
Two identical IR-UWB radar modules (NVA-R661, Novelda) were installed; one (upper radar) was aimed at the user’s lips using patch antennas, and the other (lower radar) was directed at the user’s chin using sinuous antennas with a dielectric lens. 
After testing various combinations of patch antennas and sinuous antennas for the upper and lower radars, the aforementioned configuration yielded the best performance. 

The beam width of the patch antenna is approximately 55\textdegree (vertical) $\times$ 50\textdegree (horizontal), while the sinuous antenna with a dielectric lens has a beam width of approximately 40\textdegree (vertical) $\times$ 35\textdegree (horizontal).
Therefore, the patch antenna’s wider beam pattern seems to enhance the detection of complex lip movements at close distances.
By contrast, the narrower beam pattern of the sinuous antenna with a dielectric lens below the the chin focuses on the relatively simple up-and-down movements of the chin.

To position the two radar antennas independently at locations that facilitate the acquisition of high-quality radar data, a hardware testbed was designed to incorporate the following five functionalities, as illustrated in Fig. \ref{fig:Testbed}(b): 
1) vertical sliding of the upper radar antenna, 2) vertical tilting of the upper radar antenna, 3) vertical sliding of the lower radar antenna, 4) horizontal sliding of the lower radar antenna, and 5) horizontal tilting of the lower radar antenna.

Before recruiting the participants, one of the authors of this paper conducted multiple experiments using various hardware configurations. 
Through this process, it was found that inserting a slim metal plate between the TX and RX patch antennas of the upper radar to prevent antenna coupling and incorporating a low-noise amplifier (LNA) for both radars resulted in improved performance.

\subsection{Procedure}
\label{sec:Procedure}

The participants were instructed to comfortably position themselves in front of the hardware testbed.
Once seated, the experimenter adjusted the positioning of the radar antennas to ensure that the participant's lips and tongue were within the detection range and field of view (FOV) of the upper radar, while their chin and tongue were within the detection range and FOV of the lower radar.
Specifically, the distance between the participant's lips and the antennas of the upper radar was set between 5 and 10 cm, while the distance between the participant's chin and the antennas of the lower radar was set between 10 and 15 cm.

We developed a MATLAB-based graphical user interface (GUI) to facilitate self-administered data collection.
The GUI was displayed on a desktop monitor, and the participants were instructed to pronounce each of the presented speech stimuli.
Within the GUI, the participants had the ability to independently initiate and complete the data acquisition for each speech item using clickable buttons.
Simultaneous data capture from both radars occurred when the participants pronounced the designated speech items, enabling a fair comparison of the performance between the upper and lower radars.

Applying contactless sensors in SSR presents a drawback: if the user positions the articulators outside the sensor's detection range or FOV or turns the head away from the sensor, it becomes difficult to accurately measure the articulatory movements. 
To overcome this limitation, we employed radar signals to locate the participant's articulators in a consistent position and angle prior to pronunciation.
Specifically, we established a preset position and angle, and acquired one data frame at this position and angle.
Before initiating data acquisition, the GUI window displayed a fixed radar signal representing the preset position and angle, a real-time receiving radar signal, and a correlation index indicating their similarity. 

The fixed radar signal capturing the preset position and angle, and the real-time receiving radar signal can be represented as vectors $\bm{p}$ and $\bm{q}$, respectively.
\begin{equation}
\bm{p} = [p_1, p_2, \ldots, p_{N}]
\end{equation}
\begin{equation}
\bm{q} = [q_1, q_2, \ldots, q_{N}]
\end{equation}
Here, each component in the vectors corresponds to the signal amplitude value at a corresponding fast-time index, ranging from $1$ to $N$.
Remember that $N$, representing the number of fast-time indices within a signal, is $256$ in this study.
The correlation index $\rho$ was calculated using the Pearson correlation coefficient:
\begin{equation}
\rho = \frac{\sum_{x=1}^{N} (p_{x} - \bar{p})(q_{x} - \bar{q})}{\sqrt{\sum_{x=1}^{N}(p_{x} - \bar{p})^2 \sum_{x=1}^{N}(q_{x} - \bar{q})^2}}
\end{equation}
\begin{equation}
\bar{p} = \frac{1}{N} \sum_{x=1}^{N} p_{x}
\end{equation}
\begin{equation}
\bar{q} = \frac{1}{N} \sum_{x=1}^{N} q_{x}
\end{equation}

Because the radar signal contains positional and angular information, a correlation index exceeding a certain threshold implies that the participant has positioned their articulators at a preset position and angle with acceptable tolerance. 
Once a correlation index greater than the threshold was confirmed, the participant clicked the data acquisition start button and commenced pronunciation.

The data were acquired independently for each type of speech stimulus (vowels, consonants, words, and phrases). 
Prior to data acquisition, the participants were familiarized with the pronunciation of each element within each type of speech stimulus. 
The sequential pronunciation of each item continued until all the elements of each type of stimulus were collected. 
This process was repeated 20 times, resulting in 20 sessions for each type of speech stimulus. 
Upon requests, short breaks were provided between sessions to alleviate fatigue. 
Additionally, if participants reported making a mistake while pronouncing a specific item, data acquisition for that speech item was repeated during inter-session intervals.

According to Article 15 (2) of the Bioethics and Safety Act and Article 13 of the Enforcement Rule of Bioethics and Safety Act in Korea, a research project ``which utilizes a measurement equipment with simple physical contact that does not cause any physical change in the subject'' (translated from Korean to English by the authors) is exempted from approval. 
The entire experimental procedure was designed to use only IR-UWB radars that did not cause any physical changes in the subjects.

\section{Method}
\label{sec:Method}

\subsection{Proposed Feature Extraction Algorithm}
\label{sec:FERASEC}

In this study, one frame set was acquired per radar each time a participant pronounced a speech item, as shown in Fig. \ref{fig:RawFrameset}.
We developed an algorithm to extract speech features from each frame set. 
The proposed feature extraction algorithm for each frame set can be explained as follows.

A frame set can be represented by an $M$-by-$N$ matrix, denoted as $\bm{S}$, where $M$ is the number of frames within a frame set and each row corresponds to a frame with dimensions of $1$-by-$N$.
\begin{equation}
\setlength{\arraycolsep}{0.35em}
\bm{S} = \begin{bmatrix} 
s_{1,1} & s_{1,2} & \cdots & s_{1,n} & \cdots & s_{1,N} \\ 
s_{2,1} & s_{2,2} & \cdots & s_{2,n} & \cdots & s_{2,N} \\
\vdots & \vdots   &        & \vdots  &        & \vdots  \\
s_{m,1} & s_{m,2} & \cdots & s_{m,n} & \cdots & s_{m,N} \\
\vdots & \vdots   &        & \vdots  &        &  \vdots \\
s_{M,1} & s_{M,2} & \cdots & s_{M,n} & \cdots & s_{M,N}
\end{bmatrix}
\label{eq:S}
\end{equation}
Here, $s_{m,n}$ corresponds to the amplitude of a signal at slow-time index $m$ and fast-time index $n$. 
When dealing with the clutter amplitude, received signal amplitude, and clutter-reduced signal amplitude, $s_{m,n}$ can be replaced by $c_{m}[n]$, $r_{m}[n]$, and $y_{m}[n]$, respectively.
The relationships among $c_{m}[n]$, $r_{m}[n]$, and $y_{m}[n]$ were explained by (\ref{eq:LBF_clutter}) and (\ref{eq:clutter_reduction}) in Section \ref{sec:Principles}.

Our transformation algorithm converts a two-dimensional frame set $\bm{S}$ into a one-dimensional feature sequence that effectively captures the articulatory movements of the user. 
The transformation algorithm works in the following four steps:

\begin{itemize}
\item All the frames in a given frame set (i.e., rows in $\bm{S}$) are concatenated to form a single row vector as follows. 
\begin{equation}
\setlength{\arraycolsep}{0.35em}
\begin{aligned}
\bm{f} = vec(\bm{S}) = [s_{1,1}, s_{1,2}, \ldots, s_{1,N}, s_{2,1}, \ldots, s_{M,N}]
\end{aligned}
\end{equation}
\begin{equation}
f_{i} = vec(\bm{S})_{i}
\end{equation}
Here, $vec$ represents vectorization, reshaping the matrix into a single row vector by concatenating its rows sequentially from top to bottom.
The index $i$ represents each element's position in the resulting row vector $\bm{f}$. 
The dimensions of $\bm{f}$ are $1$-by-$M N$.

\item The envelope of the concatenated frames (i.e., $\bm{f}$) is extracted. 
To achieve this, a $W$-length window is slid over the concatenated frames with a step size of one. 
The root mean square (RMS) value of the data within each window is calculated as: 
\begin{equation}
e_{j} = \sqrt{\frac{1}{W}\sum_{i=\max(1, j-\frac{W}{2})}^{\min(M N, j+\frac{W}{2}-1)} f_{i}}^2 
\end{equation}
where $j$ spans from $1$ to $M N$ and $e_{j}$ represents the magnitude of the envelope at index $j$.
The window length $W$ is set to $400$ in this study.

\item The envelope of the concatenated frames is downsampled as:
\begin{equation}
v_{k} = e_{Dk}
\end{equation}
where $k$ varies from $1$ to $\lfloor M N / D \rfloor$ and $v_{k}$ denotes the downsampled envelope value at index $k$.
The downsampling factor $D$ is set to 1024 in this study.

\item To remove the DC offset, the mean value is subtracted from each downsampled envelope value:
\begin{equation}
z_{k} = v_{k} - \bar{v}
\end{equation}
\begin{equation}
\bar{v} = \frac{1}{\lfloor M N / D \rfloor} \sum_{k=1}^{\lfloor M N / D \rfloor} v_{k}
\end{equation}
Here, $\bar{v}$ denotes the mean value of the downsampled envelope.
The resulting $z_{k}$ represents each value at index $k$ in the final one-dimensional feature sequence. 
\end{itemize}

As a result, our transformation algorithm generates a feature sequence $\bm{z}$ with dimensions of $1$-by-$\lfloor M N / D \rfloor$.
\begin{equation}
\bm{z} = [z_1, z_2, \ldots, z_k, \ldots, z_{\lfloor M N / D \rfloor}]
\end{equation}
In this study, $N$ (the number of fast-time indices in a given frame) is 256 and $D$ (the downsampling factor) is set to $1024$.
Consequently, the feature sequence $\bm{z}$ has dimensions of $1$-by-$\lfloor M/4 \rfloor$.

Two individual one-dimensional speech feature sequences are generated by applying the same transformation algorithm to the raw frame set and its clutter-reduced frame set.
They can be obtained by substituting $s_{m,n}$ in (\ref{eq:S}) with $r_{m}[n]$ and $y_{m}[n]$ in our transformation algorithm for the raw and clutter-reduced frame sets, respectively. 
These are referred to as the first and second features, respectively. 

Derivative features are then obtained.
The third and fourth features are the first derivatives of the first and second features, respectively. 
The fifth and sixth features are the second derivatives of the first and second features, respectively. 
We follow the approach of \cite{Rabiner2010} to calculate the derivatives.
The first and second derivatives are calculated using the following two equations:
\begin{equation}
\dot{z}_{k} = \frac{\sum_{l = -\lfloor L/2 \rfloor}^{\lfloor L/2 \rfloor} l \hspace{.2em} z_{k+l}}{\sum_{l=- \lfloor L/2 \rfloor}^{\lfloor L/2 \rfloor} l^2}
\label{eq:z_dot}
\end{equation}
\begin{equation}
\ddot{z}_{k} = \frac{\sum_{l=-\lfloor L/2 \rfloor}^{\lfloor L/2 \rfloor} l \hspace{.2em} \dot{z}_{k+l}}{\sum_{l=-\lfloor L/2 \rfloor}^{\lfloor L/2 \rfloor} l^2}
\label{eq:z_dotdot}
\end{equation}
where $\dot{z}_{k}$ and $\ddot{z}_{k}$ are the first and second derivative features at index $k$.
During the summation, if $k+l$ is less than $1$ or greater than $M N$, $z_{k+l}$ in (\ref{eq:z_dot}) and $\dot{z}_{k+l}$ in (\ref{eq:z_dotdot}) are treated as 0. 
We set the delta window length $L$ to $9$ in this study.

As a result, we obtain six features, each with a dimension of $1$-by-$\lfloor M/4 \rfloor$. 
Thus, the resulting dimension of the feature matrix, composed of the six speech features extracted from a single frame set, is $6$-by-$\lfloor M/4 \rfloor$.

We name this \underline{f}eature \underline{e}xtraction algorithm for IR-UWB \underline{ra}dar-based \underline{S}SR, which uses the \underline{e}nvelope of the \underline{c}oncatenated frames derived from the raw and clutter-reduced frame sets, FERASEC.
Among the six features obtained by FERASEC, the first and second features are essential, as the remaining features are delta features derived from them. 
The first and second features represent the abbreviated envelopes of the concatenated frames from the raw and clutter-reduced frame sets, respectively.

\subsection{Motivation of FERASEC}
\label{sec:FERASEC_motivation}

The motivation behind our feature extraction algorithm, which converts the frame set into an abbreviated envelope of the concatenated frames, is rooted in the inefficacy of raw IR-UWB radar data as a representation (feature) for SSR. 
This issue is discussed in detail in Section \ref{sec:FeatureNecessity}. 
The nature of IR-UWB radar data, characterized by sequential data with a large number of channels (256 channels), introduces complexity and redundancy, posing challenges for effective pattern recognition. 
To address this, our feature extraction algorithm incorporates a transformation that condenses the frame set into an abbreviated envelope of concatenated frames, reducing the channel dimension from 256 to 1. 
This design aims to extract a more efficient and effective representation of speech movement from IR-UWB radar data.

The following is the physical interpretation of the transformation algorithm.
Each frame in the set represents the normalized signal amplitudes corresponding to 256 fast-time indices that indicate the target distance from the radar. 
Therefore, the information regarding articulator movements captured by the radar is contained in the amplitude changes within the concatenated frames. 
The process of envelope detection and downsampling, used to create the abbreviated envelope, serves to emphasize the amplitude variations and reduce noise. 
Consequently, we expect that the abbreviated envelope of the concatenated frames will effectively reflect articulatory movements within the detection range and FOV of the radar. 

As necessary speech features from IR-UWB radar data, the two individually obtained abbreviated envelopes from the raw and clutter-reduced frame sets are utilized. 
The raw frame set contains articulatory movement information but is contaminated by clutter, whereas the clutter-reduced frame set loses some articulatory movement information but exhibits less clutter interference. 
Thus, we expected that utilizing both would be beneficial since they would be complementary features.
Derivative features are added to effectively capture the temporal dynamics of the proposed features. 
Derivative features are also used in the ASR field for the same purpose \cite{Chen2014}.

\begin{figure}
  \centering
  \includegraphics[width=1\linewidth]{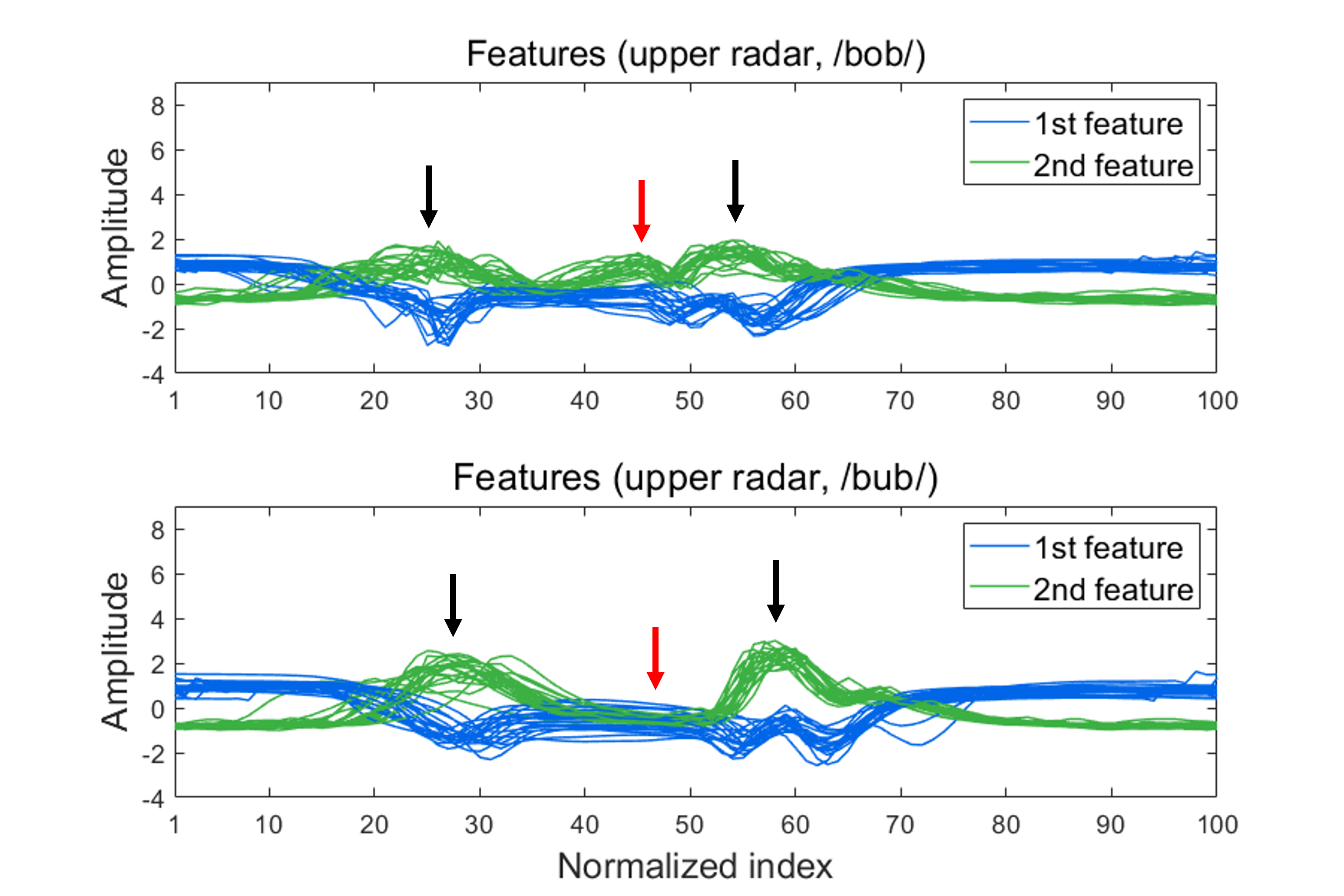}
  \caption{First and second features extracted from 20 frame sets measured by the upper radar for a participant's articulatory movements for producing /\textipa{bob}/ (top) and /\textipa{bub}/ (bottom).}
  \label{fig:Features}
\end{figure}

Fig. \ref{fig:Features} illustrates the first and second features extracted from 20 frame sets obtained by the upper radar using FERASEC. 
These frame sets captured the articulatory movements of a participant while producing two CVC syllables (/\textipa{bob}/ and /\textipa{bub}/). 
Each CVC syllable was pronounced 20 times, resulting in 20 blue curves representing the first feature and 20 green curves representing the second feature. 
For simplicity, the third through sixth features generated by FERASEC are not shown.

To compare the features extracted from each CVC syllable, the curves representing each feature in Fig. \ref{fig:Features} are aligned relative to the reference curve. 
Firstly, the size of each $1$-by-$\lfloor M/4 \rfloor$ feature sequence is normalized to 100 through interpolation if $\lfloor M/4 \rfloor$ is less than 100, or downsampling if $\lfloor M/4 \rfloor$ is greater than 100. 
The value of $M$ depends on pronunciation duration, which can vary for each utterance. 
Therefore, normalization is necessary to ensure proper comparison of features across the 20 pronunciations. 
Additionally, the onset of each pronunciation after clicking on the record button can also vary. 
Hence, one curve is selected as the reference curve, and each of the remaining 19 curves is circularly shifted until the correlation with the reference curve is maximized (i.e., until the best alignment is achieved).

In Fig. \ref{fig:Features}, the first and second features of each CVC syllable exhibit distinct patterns. 
For instance, the articulatory movements associated with the bilabial consonant (/\textipa{b}/), which appears as the first and last consonant in /\textipa{bob}/ and /\textipa{bub}/, can be observed in specific sections of the first and second features, as indicated by the black arrows. 
Notably, the extracted features of the bilabial consonant exhibit different shapes, depending on whether it is pronounced before or after the vowel. 
Moreover, the second feature, highlighted by the red arrows, reveals contrasting patterns for the two different vowels in /\textipa{bob}/ and /\textipa{bub}/.
The /\textipa{o}/ demonstrates an upward fluctuation, whereas /\textipa{u}/ shows no fluctuation. 
Therefore, by leveraging the first and second features, along with the remaining four features obtained by FERASEC (not displayed in Fig. \ref{fig:Features}), their combined utilization holds significant potential for accurately classifying various consonant and vowel pronunciations. 

\subsection{Classification Algorithms} 
\label{sec:ClassAlgo}

The length of each feature (i.e., $\lfloor M/4 \rfloor$) extracted from the radar data depends on pronunciation duration.
For example, in a three-second pronunciation where the radar acquires 200 frames per second, $M$ corresponds to $200 \times 3 = 600$ frames.
Consequently, the length of each feature becomes $\lfloor M/4 \rfloor = 150$.
Therefore, an algorithm capable of effectively classifying sequential data of varying lengths is necessary for SSR.
Various algorithms have been employed for sequence classification tasks, including dynamic time warping (DTW) \cite{Huang2002, Smith2010, Bashir2011, Jeong2011, Switonski2019}, long short-term memory (LSTM) \cite{Wang2018, Saadatnejad2019}, bidirectional long short-term memory (BLSTM) \cite{Li2020, Wagner2022:radar}, Gaussian mixture model--hidden Markov model (GMM--HMM) \cite{Reed2009, Hofe2013, Meltzner2018, Peng2019, Celler2020}, and deep neural network--hidden Markov model (DNN--HMM) \cite{Li2013, Tan2018}.

In a previous study \cite{Shin16:Towards}, a 10-word classification task was performed using the IR-UWB radar used in our study. 
Feature extraction was performed using the short-template-based CLEAN algorithm, while the classification was performed using the multidimensional dynamic time warping (MD-DTW) algorithm. 
MD-DTW can measure the distance between two multidimensional sequential data with different durations and execution speeds using nonlinear alignments \cite{Shin16:Towards}. 
Leveraging this capability, Shin and Seo \cite{Shin16:Towards} employed MD-DTW as a classification algorithm. 
The classification process involved computing and comparing the distances between the test and reference data.

To verify the effectiveness of FERASEC compared to the short-template-based CLEAN, we implemented FERASEC and MD-DTW and compared their classification performance with the baseline method of short-template-based CLEAN and MD-DTW, as summarized in Table \ref{tab:AvgClassAcc}. 
We validated the correct implementation of the baseline method by conducting the same 10-word classification task as described in \cite{Shin16:Towards}. 
Under experimental conditions that are nearly identical to those in \cite{Shin16:Towards}, we achieved a classification accuracy of 88\% using our baseline implementation. 
Considering the average classification accuracy of 84.5\% reported in \cite{Shin16:Towards}, it is evident that the baseline method was adequately implemented.
Additionally, we implemented a method that used FERASEC for feature extraction and a DNN--HMM for classification because a DNN--HMM is a representative deep learning model for handling phoneme-level speech units, such as monophones or triphones, in ASR \cite{Romdahni2015} and SSR \cite{Kim2017, Ji2018:im+US}.

\begin{figure}
  \centering
  \includegraphics[width=1\linewidth]{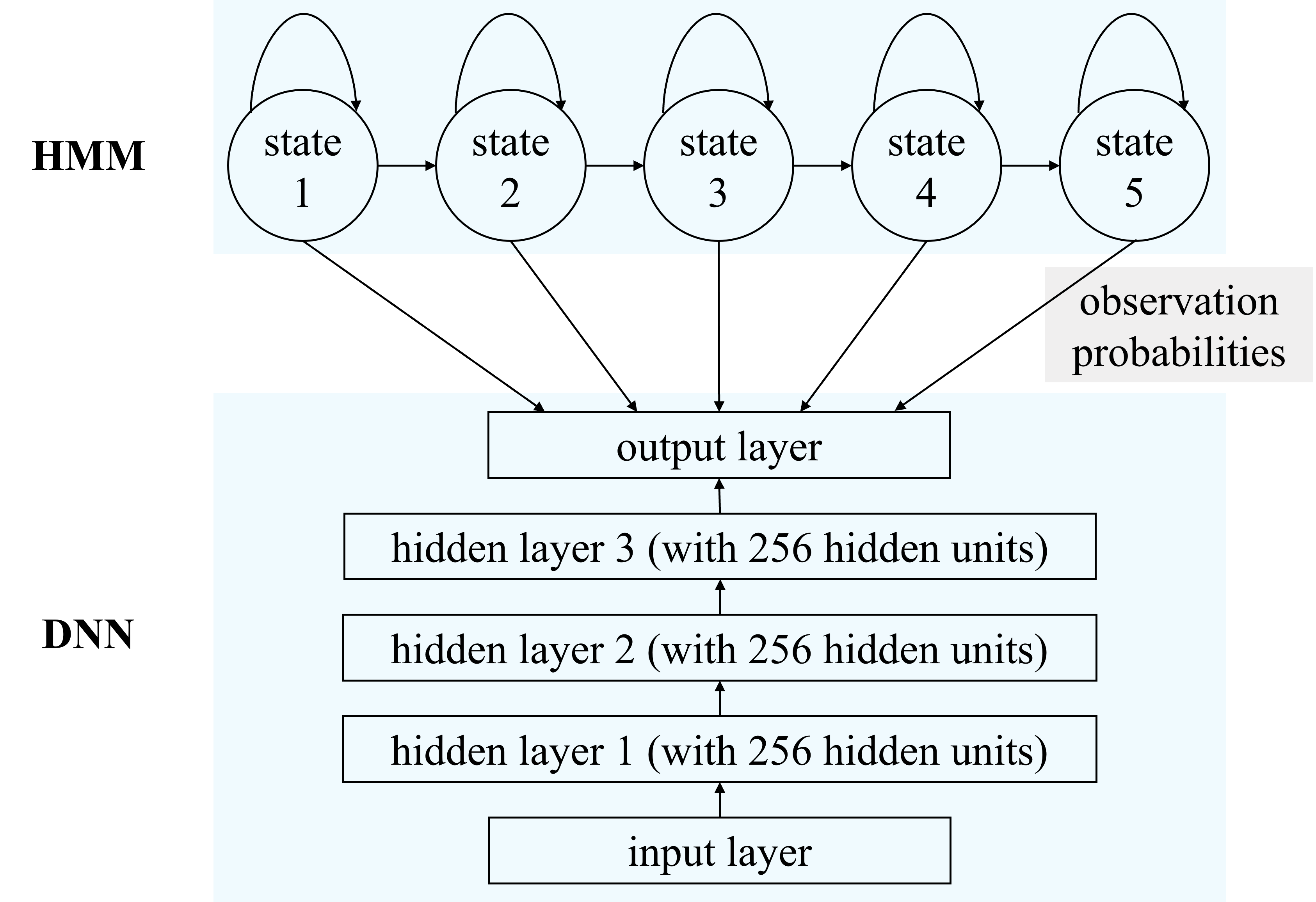}
  \caption{Structure of the DNN--HMM used in this study.}
  \label{fig:DNNHMM}
\end{figure}

In this study, we used a five-state left-to-right HMM and a DNN with three hidden layers, each consisting of 256 hidden units, to construct the DNN--HMM, as shown in Fig. \ref{fig:DNNHMM}.
The size of the ``context window'' \cite{Li2013, Romdahni2015, Murua2023}, which affects the size of the input features for the DNN, was set to seven (3-1-3).
A comprehensive description of the implementation of the DNN--HMM for the classification task can be found in \cite{Li2013}.

In total, we implemented and compared three types of methods:
\begin{itemize}
\item Short-template-based CLEAN + MD-DTW (baseline) \cite{Shin16:Towards}
\item FERASEC + MD-DTW
\item FERASEC + DNN--HMM
\end{itemize}

All the classification results in Table \ref{tab:AvgClassAcc} were obtained using leave-one-out cross-validation (LOOCV) \cite{Birkholz2018:radar}. 
Let $A$ represent the number of frame sets, and $B$ represent the number of classes for a specific speech stimulus (vowel, consonant, word, or phrase). 
For example, in our experiments, we had $A = 20 \times 8 = 160$ for vowels, as each vowel was pronounced 20 times, and there were 8 possible vowel classes. 
Similarly, we had $B = 8$ for vowels. 
LOOCV uses each of the $A$ frame sets as a test sample, and the remaining $A-1$ frame sets are used for classification.

MD-DTW calculates the distance between two frame sets based on their multidimensional speech features.
When FERASEC is applied, a multidimensional speech feature with dimension of $6$-by-$\lfloor M/4 \rfloor$ is extracted from each frame set, as described in Section \ref{sec:FERASEC}.
During the LOOCV process using MD-DTW, each test frame set is classified into the category of one of the remaining $A-1$ frame sets (i.e., labeled reference frame sets) that has the smallest distance to the test frame set calculated based on their multidimensional features.
The HMM calculates the probability of a frame set, represented by its multidimensional feature, being observed by the corresponding model.
In the case of the DNN--HMM, the multidimensional features of all frame sets, except one test frame set, are used to train the $B$ HMMs.
Each test frame set, represented by its multidimensional feature, is then classified into the category of the HMM that provides the highest probability.

\section{Results and Discussion}
\label{sec:ResultsEval}

\subsection{Performance Comparison Across Applied Methods}

\begin{table*}
\renewcommand{\arraystretch}{1.3}
\setlength{\tabcolsep}{0.8em}
\centering
\caption{Average classification accuracies (\%) for vowels, consonants, words, and phrases based on method and radar position (20 participants were involved in vowel and consonant classification, while word and phrase classification involved 4 participants).}
\label{tab:AvgClassAcc}
\begin{tabular}{cccccc} 
\hline\hline
\multicolumn{1}{c}{\multirow{1}{*}{\textbf{Method}}} & \multicolumn{1}{c}{\multirow{1}{*}{\textbf{Radar position}}} & \textbf{8 vowels} & \textbf{11 consonants} & \textbf{25 words} & \textbf{12 phrases}  \\
\hline
\multirow{2}{6cm}{\centering{Short-template-based CLEAN \\ + MD-DTW (baseline) \cite{Shin16:Towards}}} 
& Upper                   & 51.59          & 42.68         & 43.15     & 61.46         \\
& Lower                   & 40.88          & 32.13         & 19.45     & 44.69         \\
\hline                                                                             
\multirow{2}{6cm}{\centering{FERASEC + MD-DTW}}                                                                  
& Upper                   & 80.56         & 74.50        & 87.85     & \textbf{98.02}\\
& Lower                   & 66.91         & 58.37        & 73.05     & 95.62         \\
\hline     
\multirow{2}{6cm}{\centering{FERASEC + DNN--HMM}}                                                                
& Upper                   & \textbf{86.47} & \textbf{81.59} & \textbf{88.95}  & 96.88\\
& Lower                   & 70.59          & 63.57         & 81.10     & 94.27        \\
\hline\hline
\multicolumn{2}{l}{Note: The highest accuracy for each speech stimulus (column) is highlighted in bold font.}
\end{tabular}
\end{table*}

Table \ref{tab:AvgClassAcc} summarizes the classification results for the vowels, consonants, words, and phrases categorized by the applied methods and radar positions.
The average classification accuracy was computed based on the individual accuracies of the participants mentioned in Section \ref{sec:Participants}.
To calculate the classification accuracy of each participant for each speech stimulus and radar position, we used the formula ${x/(20 \times B)} \times 100$, where $x$ represents the number of frame sets accurately classified and $B$ denotes the number of classes within the specific speech stimulus type (e.g., $B = 8$ for vowels, with each vowel being pronounced 20 times).

As can be observed from Table \ref{tab:AvgClassAcc}, the average classification accuracy of FERASEC + MD-DTW consistently surpassed that of the baseline method (short-template-based CLEAN + MD-DTW) for the same speech stimulus type and radar position configuration. 
This suggests that FERASEC is more proficient than the short-template-based CLEAN in extracting speech features that accurately capture the articulatory movements from the radar data.

As explained in \cite{Shin16:Towards}, the short-template-based CLEAN algorithm is primarily designed to detect the nearest target, which may not be advantageous for extracting tongue movement information. 
Comparing with the lips or chin, the tongue is located further from the radar. 
However, capturing tongue movement information is crucial for recognizing various pronunciations. 
In contrast, FERASEC is designed to extract speech features from the entire radar measurements, encompassing all articulatory movement information, rather than solely focusing on the nearest target information. 
Consequently, FERASEC can effectively capture the movement information of the tongue, as well as the lips and chin. 
This difference in design explains why FERASEC outperforms the short-template-based CLEAN algorithm.

The performances of two different classification algorithms (MD-DTW and DNN--HMM) with the same feature extraction algorithm (FERASEC) are compared in Table \ref{tab:AvgClassAcc}.
The method that employed DNN--HMM demonstrated higher average accuracies for the vowel, consonant, and word classification tasks, whereas the method that used MD-DTW exhibited better average accuracies for the phrase classification task under the same radar position.

If we focus on the classification results with the upper radar position, which is generally better than the lower radar case, it is noteworthy that FERASEC + DNN--HMM clearly provided better classification accuracy for vowels and consonants than that obtained by FERASEC + MD-DTW.
For the relatively easier classification tasks of words and phrases, both methods demonstrated similar performance.

\subsection{Performance Comparison Across Radar Positions}

As presented in Table \ref{tab:AvgClassAcc}, the classification accuracy consistently improved when the radar was positioned at the upper location compared to the lower location, regardless of the applied methods and types of speech stimuli.
Considering that the short-template-based CLEAN algorithm is specialized for extracting the movement information of the nearest target, the upper radar position is advantageous for capturing the movement information of the lips, whereas the lower position is advantageous for capturing the movement information of the chin.
Therefore, the higher classification accuracies obtained with the upper radar position using the short-template-based CLEAN algorithm, compared with those achieved with the lower position, suggest that the movement information of the lips plays a more crucial role than that of the chin in recognizing silent speech when the movement information of the tongue cannot be obtained.

FERASEC was designed to extract information on all articulatory movements, including tongue motion, from radar measurements. 
We observed that the received radar signals changed in response to tongue motion when the radar was positioned in front of the lips or below the chin. 
However, when the radar was positioned in front of the lips, the change in the radar signals was not clearly noticeable when they were occluded by the teeth during certain pronunciations. 
For instance, the radar signals showed distinct changes according to tongue motion when the mouth was open and the upper and lower teeth did not block the radar signals (as in pronouncing /\textscripta/); however, the radar signals hardly changed when the upper and lower teeth blocked the signals (as in pronouncing /\textipa{i}/). 
This finding supports the observation that certain consonants are difficult to distinguish because of the presence of upper teeth in the upper radar configuration, as discussed in Section \ref{sec:ConsonantClassification}. 
The signal blockage by the teeth when obtaining tongue motion information is not an issue when the radar is positioned below the chin. 
However, in this case, it becomes challenging to obtain information about lip motions.

In summary, when using FERASEC, the upper radar position is beneficial for capturing lip motion information; however, it may result in the loss of some tongue motion information during certain pronunciations. 
In contrast, the lower radar position is advantageous for capturing chin and tongue motion information; however, it may partially lose lip motion information. 
The superior performance of the upper radar position, as shown in Table \ref{tab:AvgClassAcc}, confirms that detecting lip motions rather than chin motions is crucial for recognizing diverse pronunciations, although this may involve the loss of some tongue movement information.

\subsection{Confusion Matrix Analysis}

\begin{figure*}
  \centering
  \includegraphics[trim={0cm 0cm 0cm 0.1cm}, width=1\linewidth]{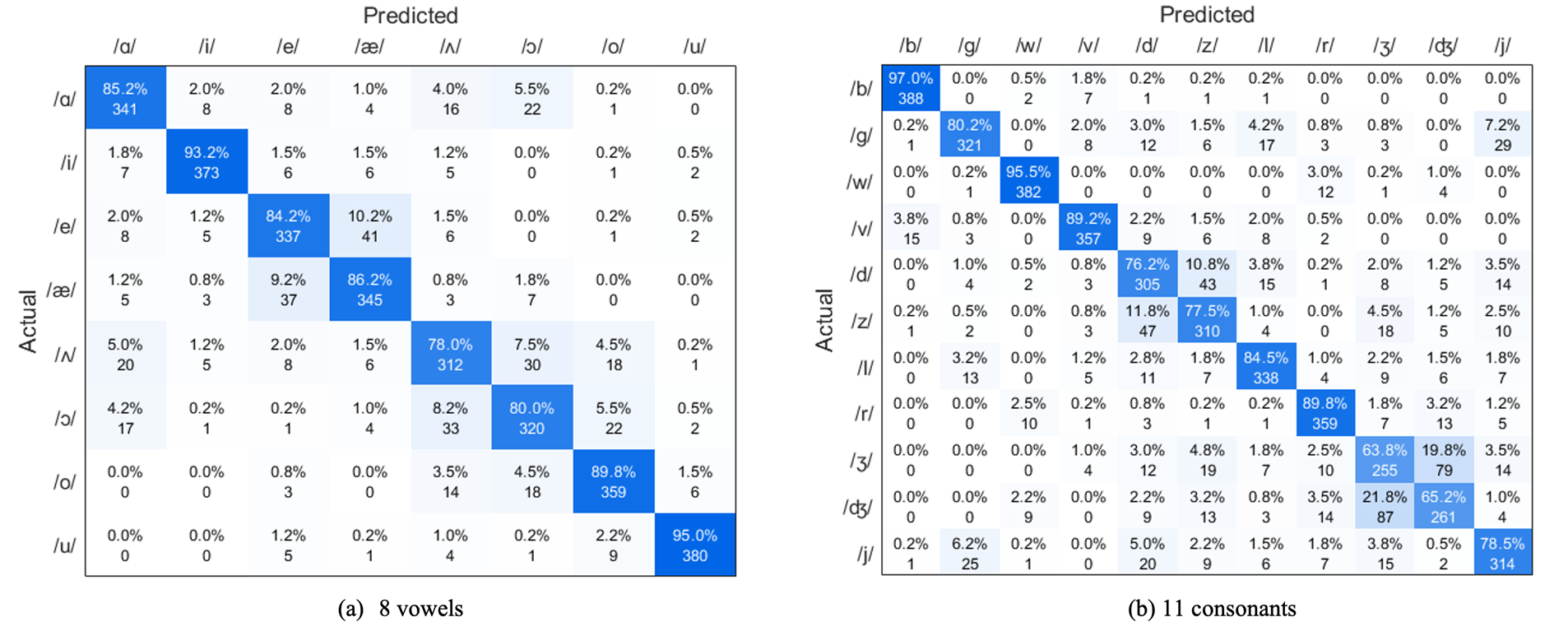}
  \caption{Confusion matrices across 20 participants for (a) vowel and (b) consonant classification tasks when the radar was positioned in front of the lips. The FERASEC + DNN--HMM method was used.}
  \label{fig:ConfMat}
\end{figure*}

We generated confusion matrices for the vowel and consonant classification tasks using the upper radar configuration and FERASEC + DNN--HMM method, as shown in Fig. \ref{fig:ConfMat}.
Confusion matrices for vowel and consonant classification tasks are included because they provide a more direct analysis of articulator movements at the phonemic level than word- or phrase-level tasks. 
Each element of the matrix contains the relative accuracy (\%) and the number of data samples (frame sets) of the actual pronounced phonemes (rows) predicted as a specific phoneme (columns).
For example, in Fig. \ref{fig:ConfMat}(a), from 400 data samples of pronouncing /\textipa{i}/, 373 samples (i.e., 93.2\%) were correctly predicted, while 6 samples (i.e., 1.5\%) were incorrectly predicted as /\textipa{\ae}/.
In each row, the sum of the relative accuracies and number of data samples are 100\% and 400, respectively (it should be noted that each of the 20 participants pronounced each phoneme 20 times).

Before analyzing the results based on the confusion matrix in Fig. \ref{fig:ConfMat}, it is informative to introduce a previous study \cite{Wang2013:EMA} that examined the articulatory distinctiveness of vowels and consonants using EMA sensors attached to the lips and tongue.
The vowels and consonants used as speech stimuli in \cite{Wang2013:EMA} are identical to those used in our study.

Wang \textit{et al.} \cite{Wang2013:EMA} observed that higher classification accuracy was achieved for high and front vowels (/\textipa{i}/, /\textipa{e}/, /\textipa{\ae}/, and /\textipa{u}/) than that obtained for low and back vowels (/\textscripta/, /\textturnv/, /\textopeno/, and /\textipa{o}/).
This distinction arises from differences in tongue position during articulation, with high and front vowels produced with a high and front tongue position, whereas low and back vowels involve a low and back tongue position.
In the consonant classification task, the most frequent confusion occurred between /\textyogh/ and /\textdyoghlig/, which require relatively similar articulation places.

\subsubsection{Vowel Classification} 

As observed in Fig. \ref{fig:ConfMat}(a), the upper radar-based vowel classification task demonstrated higher classification accuracies for /\textipa{i}/, /\textipa{\ae}/, /\textipa{o}/, and /\textipa{u}/ compared to that for /\textscripta/, /\textipa{e}/, /\textturnv/, and /\textopeno/.
This result partially aligns with the findings of the earlier EMA sensor-based study \cite{Wang2013:EMA} in which the high and front vowels /\textipa{i}/, /\textipa{\ae}/, and /\textipa{u}/ were distinguished more easily than the low and back vowels /\textscripta/, /\textturnv/, and /\textopeno/.
However, it is noteworthy that the high and front vowel /\textipa{e}/ exhibited relatively lower accuracies, whereas the low and back vowel /\textipa{o}/ demonstrated relatively higher accuracies in our radar-based study.

The classification errors for the high and front vowel /\textipa{e}/ primarily stem from confusion with another high and front vowel /\textipa{\ae}/.
Compared with the use of multiple EMA sensors directly attached to the lips and tongue, the contactless IR-UWB radar employed in SSR may be less effective in capturing subtle tongue movements.
However, the radar appears to be more efficient in detecting fine lip movements by capturing the movements of the surrounding muscles.
Thus, vowels that are relatively close in the ``vowel space'' \cite{Wang2013:EMA} (i.e., vowels that exhibit similar tongue movements) and require relatively similar lip movements during pronunciation may pose a greater challenge for distinction.
This could explain the frequent confusion observed between /\textipa{e}/ and /\textipa{\ae}/.
By contrast, vowels that are relatively close in the vowel space but require distinct lip movements can be classified with less confusion, as evidenced by the distinction between /\textipa{i}/ and /\textipa{e}/.

The relatively high classification accuracy of the low and back vowel /\textipa{o}/ can be attributed to its requirement of distinct lip protrusion, which can be effectively detected by the upper IR-UWB radar.

\subsubsection{Consonant Classification} 
\label{sec:ConsonantClassification}

From Fig. \ref{fig:ConfMat}(b), it can be observed that the consonants /\textipa{b}/, /\textipa{w}/, /\textipa{v}/, and /\textipa{r}/ achieved higher classification accuracies than the other consonants in the upper radar configuration.
This can be attributed to the fact that these consonants are pronounced with distinct lip movements that can be effectively captured by the upper IR-UWB radar.
Specifically, /\textipa{b}/ and /\textipa{w}/ involve distinguishable movements of both lips, /\textipa{v}/ is a labiodental consonant produced with the lower lip contacting the upper teeth, and /\textipa{r}/ requires lip protrusion during production \cite{King2020}.

Furthermore, it is noteworthy that significant confusions occurred between /\textyogh/ and /\textdyoghlig/, as well as between /\textipa{d}/ and /\textipa{z}/.
The confusion between /\textyogh/ and /\textdyoghlig/ is primarily attributed to their similar places of articulation, as observed in the previously mentioned EMA sensor-based study \cite{Wang2013:EMA}.
Although /\textipa{d}/ and /\textipa{z}/ were not reported to be frequently confused in \cite{Wang2013:EMA}, both are alveolar sounds that require either the tongue tip or blade to touch the alveolar ridge during pronunciation, which is a challenging area for accurate measurement using the upper radar.
When the IR-UWB radar is positioned in front of the lips, the classification of alveolar sounds may be limited owing to the reduced penetrability of the IR-UWB radar signal caused by the upper teeth.

\subsection{Performance Comparison with Other Contactless Radar-Based SSR Studies}

In this section, we compare the performance of our method with that of other contactless radar-based SSR methods reported in the literature. 
Our study achieved average classification accuracies of 86.47\%, 81.59\%, 88.95\%, and 96.88\% for the vowels, consonants, words, and phrases, respectively.
These results were obtained using the FERASEC + DNN--HMM method with the upper radar position.

Shin and Seo \cite{Shin16:Towards} achieved accuracies of 94\% and 84.5\% for the 5-vowel (/\textipa{\textscripta}/, /\textipa{\ae}/, /\textipa{i}/, /\textopeno/, and /\textipa{u}/) and 10-word (zero to nine) classification tasks, respectively, with five speakers.
While their accuracy for vowel classification (94\%) is higher than ours (86.47\%), it is important to note that their vowel corpus size is smaller than ours, and the five vowels they used are highly distinguishable.
As indicated in Table \ref{tab:AvgClassAcc}, with the same vowel corpus as that used in our study, the method proposed in \cite{Shin16:Towards} achieved an accuracy of only 51.59\%, whereas our method (FERASEC + DNN--HMM) achieved a significantly higher accuracy of 86.47\%.
Ferreira \textit{et al.} \cite{Ferreira2022:radar} reported an accuracy of 88.3\% for a 13-word European Portuguese classification task with four participants using an FMCW radar.

The word corpus size in our study is larger than those in the previously mentioned contactless radar-based SSR studies.
Furthermore, our word corpus is designed to be phonetically balanced \cite{Wang2016:EMA}, in contrast to those in \cite{Shin16:Towards} and \cite{Ferreira2022:radar}.
Nevertheless, we achieved similar or higher levels of accuracy in word classification tasks than those obtained in \cite{Shin16:Towards} and \cite{Ferreira2022:radar}.
We compared our method's vowel classification performance with \cite{Shin16:Towards} and its word classification performance with both \cite{Shin16:Towards} and \cite{Ferreira2022:radar}.
While our method is capable of classifying vowels, consonants, words, and phrases, it is noteworthy that \cite{Shin16:Towards} and \cite{Ferreira2022:radar} did not specifically address the recognition of speech units other than vowels or words.

Recently, Zeng \textit{et al.} \cite{Zeng2023} reported successful recognition of individual words within 1000 everyday conversation sentences using an FMCW radar. 
While their achievement in word recognition at the sentence level is notable, direct performance comparison with our results is not possible because the types of SSR tasks are different. 
The internal language model in their work adds complexity to such a comparison. 
Our SSR task relied solely on articulatory movement information, while in \cite{Zeng2023}, it involved both articulatory movement and context information from an internal language model.
The internal language model enhances word recognition performance by leveraging relationships among the listed words. 

Zeng \textit{et al.} \cite{Zeng2023} did not present phoneme-level recognition results or analysis, whereas our main contribution is phoneme-level SSR. 
As mentioned in Section \ref{sec:Intro}, the IR-UWB radar used in our study has higher performance potential than conventional radars. 
We demonstrated the first contactless radar-based SSR of phonemes in this study. 
The capability of phoneme recognition can be extended to the recognition of diverse speech, composed of a set of many phonemes.

\subsection{Discussion}

\subsubsection{Necessity of Developing a Feature Extraction Algorithm for IR-UWB Radar-Based SSR}
\label{sec:FeatureNecessity}

\begin{table}
\renewcommand{\arraystretch}{1.3}
\setlength{\tabcolsep}{1.5em}
\centering
\caption{Vowel and consonant classification results when using either raw or clutter-reduced frame sets as input for DNN--HMM.}
\label{tab:RCFramePf}
\begin{tabular}{ccc}
\hline \hline
\multirow{2}{*}{\textbf{Input}} & \multicolumn{2}{c}{\textbf{Accuracy (\%)}} \\
\cline{2-3} 
& \textbf{8 vowels} & \textbf{11 consonants} \\ \hline
Raw frame set & 44.38 & 38.64 \\ \hline
Clutter reduced frame set & 48.13 & 40.45 \\ \hline \hline
\end{tabular}
\end{table}

Before developing the proposed feature extraction algorithm (i.e., FERASEC), we attempted end-to-end deep learning, a method that does not rely on explicitly engineered features, for vowel and consonant classification tasks.
Since the DNN--HMM harnesses the capabilities of deep neural networks to extract intricate patterns from the input data, it is applicable to a raw radar data-based end-to-end approach. 
Using upper radar data, we tested two different scenarios of end-to-end deep learning-based classification: employing either the raw frame set or the clutter-reduced frame set as input for DNN--HMM.

The classification results of these end-to-end approaches are summarized in Table \ref{tab:RCFramePf}. 
When the raw frame set was employed, vowel and consonant classification accuracies were 44.38\% and 38.64\%, respectively. 
When the clutter-reduced frame set was used, vowel and consonant classification accuracies improved to 48.13\% and 40.45\%, respectively. 
While a slight enhancement occurred when the clutter was mitigated from the raw IR-UWB radar data, the accuracies still did not surpass 50\% in both vowel and consonant classifications. 
This highlights that raw or clutter-reduced frame sets themselves do not provide an effective representation in IR-UWB radar-based SSR.

Although end-to-end deep learning approaches or neural network-based feature extractors, which operate without the need for explicitly engineered features, have gained popularity, explicitly engineered features continue to be widely utilized as inputs to deep learning models in various fields, owing to their compactness, discriminative power, and robustness to noise. 
For instance, in the ASR field, log mel spectrograms are still commonly employed as features instead of raw audio data. 
Likewise, the features obtained by FERASEC have the potential to serve as foundational features in future IR-UWB radar-based SSR.

\subsubsection{IR-UWB Radar-Based Contactless SSR for Future Smart Devices} 
\label{sec:FutureSSR}

In Section \ref{sec:Intro}, we highlighted that the usability of a contactless SSR system surpasses that of a contact SSR system requiring a helmet or a skin-attached antenna. 
Although the hardware testbed in Fig. \ref{fig:Testbed} may appear inconvenient for daily use, contemporary silicon die packaging technologies enable the integration of radar antennas into a chip package \cite{Jaime2016}. 
Moreover, the complete radar functionality, including the signal transceiver and antennas, can be implemented on a single chip \cite{Nasr2016}. 
Thus, radar technologies can now be deployed in commercially available space-constrained devices. 
For instance, Google's Pixel 4 smartphone incorporates a tiny single-chip radar for micro gesture recognition.

\begin{figure}
\centering
\centerline{\includegraphics[width=0.9\linewidth]{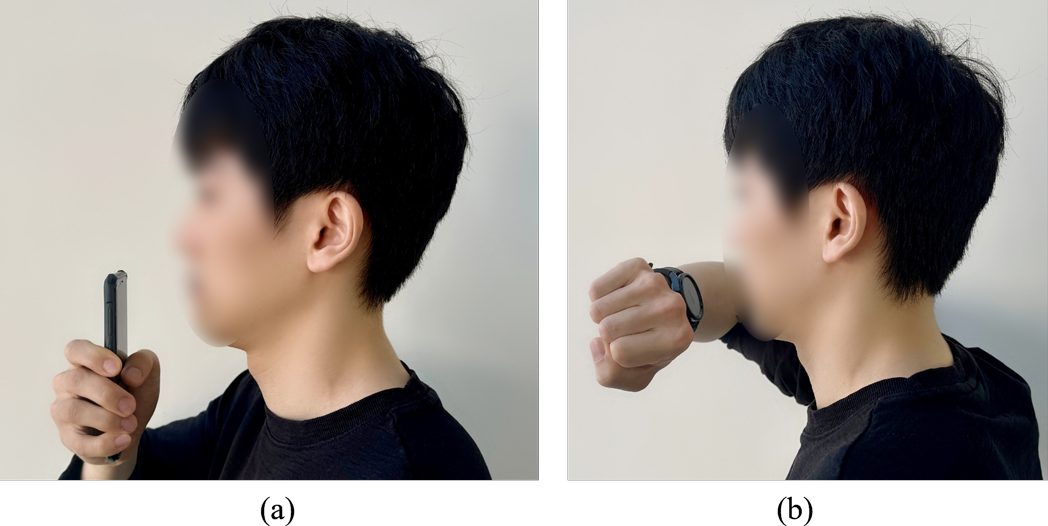}}
\caption{Potential use cases for a future (a) smartphone and (b) smartwatch with IR-UWB radar-based SSR technology.}
\label{fig:Usage}
\end{figure}

The IR-UWB radar modules used in this study are based on a single CMOS transceiver chip \cite{Yin2019}. 
With chip packaging techniques, the entire functionality of the IR-UWB radar can be integrated into a single chip. 
Thus, we anticipate that IR-UWB radar-based SSR can be deployed in commercial devices such as smartphones and smartwatches. 
Since our study suggests a single radar placed in front of the lips for IR-UWB radar-based SSR, potential use cases are illustrated in Fig. \ref{fig:Usage}.

Although contactless sensors are desirable for improving usability, the performance of SSR could degrade if the articulators are placed outside the sensor's detection range or if the articulators are not aligned with the sensor's FOV. 
To overcome this challenge, we developed an aiding algorithm to check if the position and angle between the articulators and radar sensor are proper, as explained in Section \ref{sec:Procedure}. 
This approach is directly applicable to the use cases in Fig. \ref{fig:Usage}. 
Users should adjust the distance and angle between the mouth and the smart device and commence silent speech only when the green light from the aiding algorithm is on. 
This approach enables high-quality articulatory measurements while minimizing concerns about detection range or angle variability.

In the experiment using the testbed in Fig. \ref{fig:Testbed}, we operated the algorithm of aiding a participant to find the preset position and angle based on both upper and lower radar signals. 
However, for the use cases in Fig. \ref{fig:Usage}, the aiding algorithm should be operated based on the upper radar signal alone. 
To compare the performance between the double radar and single radar-based aiding cases, one participant conducted additional vowel and consonant classification experiments in both scenarios. 
The FERASEC + DNN--HMM method was applied for classification. 
The classification accuracies for vowels and consonants were 89.38\% and 87.73\%, respectively, when both upper and lower radars were used for the aiding algorithm. 
The vowel and consonant classification accuracies slightly degraded to 88.13\% and 86.36\%, respectively, when only the upper radar was used for the aiding algorithm. 
This slight performance degradation indicates that the exclusion of the lower radar is not critical for the aiding purpose. 
Therefore, we can still use the aiding algorithm in Section \ref{sec:Procedure} for the use cases of Fig. \ref{fig:Usage}.

\section{Conclusion}
\label{sec:Conclusion}

Radar holds promise as a sensor for contactless silent speech recognition; however, the recognition of phonemes, which includes both vowels and consonants, remains a critical milestone yet to be demonstrated using contactless radars. 
The recognition of phonemes, the fundamental units of speech, is vital as it establishes the basis for recognizing diverse speech.
Phoneme recognition presents a greater challenge than word or phrase recognition because of the need to detect subtle and diverse articulatory movements that occur within very short durations. 
In this study, we successfully demonstrated the feasibility of phoneme recognition using a contactless IR-UWB radar. 
To accomplish this, we introduced a novel feature extraction algorithm called FERASEC, which effectively extracts speech features from raw radar data. 
During the development of FERASEC, our focus was on capturing movement information from all detectable articulators using the IR-UWB radar. 
We combined FERASEC with either MD-DTW or DNN--HMM for classification and evaluated its performance using two radar positions: in front of the lips or below the chin. 
The classification accuracies achieved for vowels and consonants using the FERASEC + DNN--HMM method with a radar in front of the lips provide compelling evidence of the phoneme recognition capability of IR-UWB radar-based contactless SSR technology.

\bibliographystyle{IEEEtran}
\bibliography{mybibfile, IUS_publications}

\phantomsection

\begin{IEEEbiography}[{\includegraphics[width=1in,height=1.25in,clip,keepaspectratio]{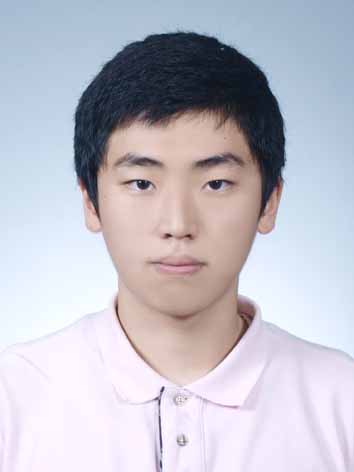}}]{Sunghwa Lee} received the B.S degree in integrated technology from Yonsei University, Incheon, Korea, in 2016. He is currently pursuing the Ph.D. degree in integrated technology at Yonsei University, Incheon, Korea. His research interests include signal processing, (silent) speech recognition, and machine learning. He was a recipient of the Undergraduate and Graduate Fellowships from the ICT Consilience Creative Program supported by the Ministry of Science and ICT, Korea.
\end{IEEEbiography}

\begin{IEEEbiography}[{\includegraphics[width=1in,height=1.25in,clip,keepaspectratio]{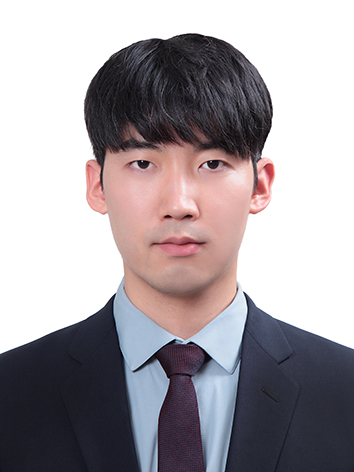}}]{Younghoon Shin} received his B.S. degree in electrical and electronic engineering from Yonsei University, Seoul, Korea, in 2013, and his Ph.D. degree in integrated technology from the same university in 2018. He contributed to this study on contactless silent speech recognition as a postdoctoral scholar at Yonsei University. Currently, he is a senior research engineer at the Robotics Lab in Hyundai Motor Company. His research interests include machine learning, computer vision, and robotics software.
\end{IEEEbiography}

\begin{IEEEbiography}[{\includegraphics[width=1in,height=1.25in,clip,keepaspectratio]{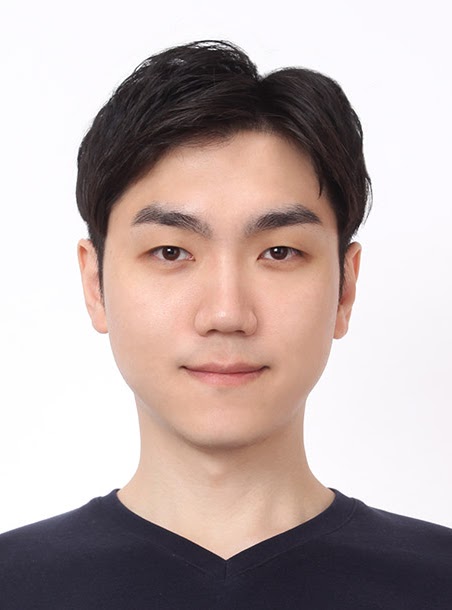}}]{Myungjong Kim} received the B.S. degree from the Department of Electronics Engineering, Tech University of Korea, Siheung, South Korea, in 2008, the M.S. degree from the Department of Information and Communications Engineering, Korea Advanced Institute of Science and Technology (KAIST), Daejeon, South Korea, in 2010, and the Ph.D. degree from the School of Electrical Engineering, KAIST, in 2016. He is currently a Deep Learning Scientist at NVIDIA Corporation, Santa Clara, CA, USA. His research interests include deep learning and signal processing for automatic speech and speaker recognition.
\end{IEEEbiography}

\begin{IEEEbiography}[{\includegraphics[width=1in,height=1.25in,clip,keepaspectratio]{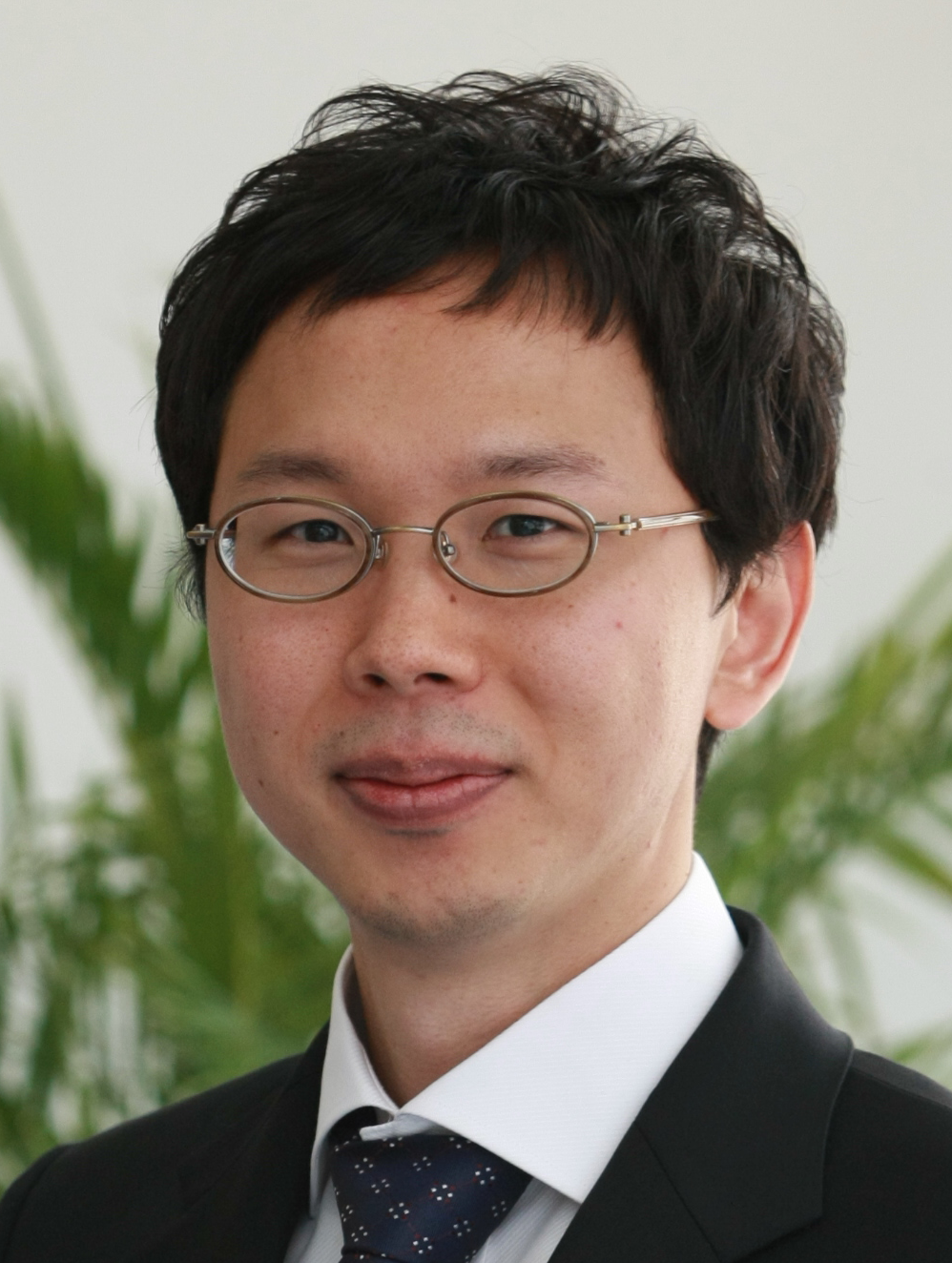}}]{Jiwon Seo} (M'13) received the B.S. degree in mechanical engineering (division of aerospace engineering) in 2002 from Korea Advanced Institute of Science and Technology (KAIST), Daejeon, Korea, and the M.S. degree in aeronautics and astronautics in 2004, the M.S. degree in electrical engineering in 2008, and the Ph.D. degree in aeronautics and astronautics in 2010 from Stanford University, Stanford, CA, USA. 
He is currently an associate professor with the School of Integrated Technology, Yonsei University, Incheon, Korea. He is also an adjunct professor with the Department of Convergence IT Engineering, Pohang University of Science and Technology (POSTECH), Pohang, Korea. His research interests include GNSS anti-jamming technologies, complementary PNT systems, and intelligent unmanned systems. 
Prof. Seo is a member of the International Advisory Council of the Resilient Navigation and Timing Foundation, Alexandria, VA, USA, and a member of the Advisory Committee on Defense of the Presidential Advisory Council on Science and Technology, Korea.
\end{IEEEbiography}

\EOD

\vfill

\end{document}